\documentclass[nofootinbib,prd,twocolumn,showpacs,amsmath,amssymb]{revtex4-1}
\usepackage{bm}
\usepackage{graphicx}
\usepackage{hyperref}

\newcommand{\rmd}{\mathrm{d}}
\newcommand{\rmi}{\mathrm{i}}
\newcommand{\rmc}{\mathrm{c}}
\newcommand{\rms}{\mathrm{s}}
\newcommand{\km}{\mathrm{km}}

\newcommand{\td}[1]{\frac{\rmd}{\rmd #1}}
\newcommand{\tdf}[2]{\frac{\rmd #2}{\rmd #1}}
\newcommand{\avg}[1]{\left\langle #1 \right\rangle}

\newcommand{\wc}{\omega_\text{c}}

\newcommand{\Gf}{G_\mathrm{F}}
\newcommand{\tomega}{\tilde\omega}
\newcommand{\tmu}{\tilde\mu}
\newcommand{\tg}{\tilde{g}}
\newcommand{\dN}{\dot{N}}

\newcommand{\bnabla}{\boldsymbol{\nabla}}

\newcommand{\sfH}{\mathsf{H}} 
\newcommand{\bfp}{\mathbf{p}} 
\newcommand{\bfx}{\mathbf{x}} 
\newcommand{\bfv}{\mathbf{v}} 
\newcommand{\bfK}{\mathbf{K}} 

\newcommand{\nue}{\protect{\nu_e}}
\newcommand{\nux}{\protect{\nu_x}}
\newcommand{\nueb}{\protect{\bar\nu_e}}
\newcommand{\nuxb}{\protect{\bar\nu_x}}

\begin{document}

\title{Fast neutrino flavor conversion: roles of dense matter and
  spectrum crossing} 

\author{Sajad Abbar}
\email{abbar@apc.in2p3.fr}
\altaffiliation{Permenant address: Astro-Particule et Cosmologie (APC),
  Université Denis Diderot, France}
\author{Huaiyu Duan}
\email{duan@unm.edu}
\affiliation{Department of Physics \& Astronomy, University of New
  Mexico, Albuquerque, NM 87131, USA}

\begin{abstract}
  The flavor conversion of a neutrino usually occurs at densities
  $\lesssim \Gf^{-1} \omega$, whether in ordinary matter or a dense
  neutrino medium, and on time/distance scales of order $\omega^{-1}$, where
  $\Gf$ is the Fermi weak coupling constant and
  $\omega$ is the typical vacuum oscillation frequency of the neutrino.
  In contrast, fast neutrino flavor conversions or fast neutrino
  oscillations can occur on
  scales much shorter than $\omega^{-1}$ in a very dense, anisotropic
  neutrino gas such as that in a core-collapse supernova or a binary
  neutron star merger. The origin of fast neutrino oscillations still seems
  elusive 
  except that it is a mathematical solution to the equation of motion.
  It has been suggested that the fast oscillations in stationary
  neutrino gases
  require a crossing in 
  the electron lepton number angular distribution of the neutrino
  and that they are suppressed at large matter densities as normal
  oscillations are. By inspecting a simple four-beam neutrino model we
  illustrate how the multi-angle effects that were once found to
  suppress collective neutrino oscillations now give rise to fast flavor
  conversions. As a result,  a large matter density can 
  induce fast oscillations in certain astrophysical scenarios such as
  at the early epoch of a core-collapse supernova. We also provide an
  explicit proof of the necessity of a crossed neutrino angular
  distribution for fast oscillations to occur in an outward flowing, axially
  symmetric  neutrino flux such as in the multi-bulb supernova
  model. However, fast oscillations can occur
  without a crossed 
  angular distribution when both inward and outward flowing neutrino
  fluxes are present in a stationary neutrino gas. 
\end{abstract}


\maketitle

\section{Introduction} 
Neutrinos are copiously produced in
core-collapse supernovae and binary neutron star mergers and play
important roles in these astrophysical events.
Because neutrinos can experience flavor
transformation or oscillations during 
propagation (see, e.g., Ref.~\cite{Patrignani:2016xqp} for a review),
it is of great interest to know whether neutrino oscillations can have
any signficant impact on the chemical evolution or even the dynamics of
these environments.

In vacuum the flavor conversion of a neutrino occurs on a length
scale determined by its vacuum 
oscillation frequency $\omega=\delta m^2/2E$, where
$\delta m^2$ and $E$ are the (effective) mass-squared difference and the
energy of the neutrino, respectively. For a 10 MeV neutrino and the
atmospheric mass splitting  the flavor conversion
occurs on the scale of 1 km. A new length scale
$\lambda^{-1}=(\sqrt2\Gf n_e)^{-1}$ comes into the play when the neutrino
propagates through a dense matter such as inside the sun, where
$\Gf$ is the Fermi weak 
coupling constant, 
and $n_e$ is the net electron number density. However,
the neutrino flavor conversion is suppressed in dense matter until
$\lambda\lesssim \omega$ where
the Mikheyev-Smirnov-Wolfenstein (MSW) mechanism can operate
\cite{Mikheyev:1985aa}. As a result, the length
scale of the neutrino flavor conversion through the MSW mechanism is
also of order $\omega^{-1}$. 

As early as in 2005 Sawyer has envisioned that flavor
conversions can occur on the scale of $(\sqrt2\Gf n_\nu)^{-1}$
inside a core-collapse supernova, where $n_\nu$ is the number density
of the neutrinos \cite{Sawyer:2005jk,Sawyer:2008zs}. It is expected
that this fast flavor conversion or fast (neutrino) oscillations can
take place over the distance of 
$0.1-1$ cm. However, such fast oscillations were not observed in
more sophisticated numerical simulations (see, e.g.,
Refs.~\cite{Duan:2006an,EstebanPretel:2007ec,Fogli:2007bk,
Duan:2010bf,Mirizzi:2011tu,Mirizzi:2012wp}; see 
also Ref.~\cite{Duan:2010bg} for a review). 
It turns out that all these calculations except those in
Refs.~\cite{Mirizzi:2011tu,Mirizzi:2012wp} adopted the neutrino bulb model
where the neutrinos of different flavors are emitted isotropically from the same
neutrino sphere. Recently Sawyer reported fast
flavor conversions in a multi-bulb model where
different neutrino flavors have distinct neutrino spheres
\cite{Sawyer:2015dsa}.  This
finding seems to have been confirmed by other groups
\cite{Chakraborty:2016lct,Dasgupta:2016dbv}. 

Although a few works have been put out with the aims
to understand fast neutrino flavor conversions from the perspective of
the dispersion 
relation \cite{Izaguirre:2016gsx,Capozzi:2017gqd} and in the nonlinear
regime \cite{Dasgupta:2017oko}, the origin of fast oscillations remains
largely elusive. In investigating  a multi-bulb model with uniform
neutrino angular distributions (on the surface of the neutrino spheres),
Dasgupta et al observed  that ``a necessary condition
to have fast instabilities \emph{appear} to be a crossing between the
angular spectra of $\nu_e$ and
$\bar{\nu}_e$'' \cite{Dasgupta:2016dbv}. However, this finding was
not conclusive because no closed-form analytical solution was
found.
They have also concluded that ``background
matter suppresses instabilities in 
space (but not in time)''.
The main goals of this paper are to understand the origin of fast oscillations
and to re-examine the above-mentioned two
conclusions about fast oscillations of
Ref.~\cite{Dasgupta:2016dbv}.

The rest of the paper is organized as follows. After establishing the
formalism in Sec.~\ref{sec:eom},  we will revisit the bipolar model
in Sec.~\ref{sec:homo} from the
perspective of spectrum crossing. 
In Sec.~\ref{sec:aniso} we will explain the origin of fast
oscillations in a four-beam neutrino 
model through its connection to the bipolar model. We will also give
an explicit proof of the necessity of a crossed neutrino angular
distribution for fast oscillations to occur in an outward flowing,
axially symmetric neutrino flux.
In Sec.~\ref{sec:2bulb} we will re-examine several important results
about fast neutrino flavor conversions in the literature that are
relevant to supernova physics. Finally, we will 
discuss the implications of our work
and conclude in
Sec.~\ref{sec:conclusions}.

\section{Equation of motion%
\label{sec:eom}}

We consider the mixing between two neutrinos
flavors, $e$ and $x$, where $x$ stands for an appropriate linear
combination of the $\mu$ and $\tau$ flavors.
We adopt the neutrino-flavor-isospin convention \cite{Duan:2005cp} in which
the neutrino and antineutrino of energy $E$ are labeled by their oscillation
frequencies
\begin{align}
\omega = \pm \frac{\delta m^2}{2 E},
\label{eq:w}
\end{align}
respectively,
where the plus (minus) sign applies to the neutrino (antineutrino), and
a positive (negative) neutrino mass-squared difference
$\delta m^2$ indicates the normal (inverted) neutrino mass hierarchy.

We assume that the physical conditions including the
densities of all neutrino species are essentially constant and
homogeneous on the time and distance scales of interest. 
Following Refs.~\cite{Dasgupta:2009mg,Izaguirre:2016gsx} we define the
electron lepton number (ELN) distribution or spectrum as
\begin{align}
g(\omega,\bfv) \propto E^2 \left|\frac{\rmd E}{\rmd\omega}\right|\times 
\begin{cases}
f_\nue(\bfp) - f_\nux(\bfp) &\text{for neutrino},\\
f_\nuxb(\bfp) - f_\nueb(\bfp) &\text{for antineutrino},
\end{cases}
\label{eq:eln}
\end{align}
where $\bfv$ and $\bfp$ are the velocity and momentum of the neutrino,
respectively, and $f_{\nu_\alpha/\bar\nu_\alpha}$ ($\alpha=e,x$)
are the occupation
numbers of the corresponding neutrino species. Here we have assumed
that neutrinos are relativistic so that $|\bfv|=1$ and $|\bfp|=E$.
There exist multiple conventions in the literature for the
normalization of the ELN spectrum. Here we choose to normalize it
by the $\nue$ density such that
\begin{align}
\frac{n_\nue-n_\nux}{n_\nue}
=\begin{cases}
\displaystyle\int\!\frac{\rmd\bfv}{4\pi}
\int_{0}^\infty\rmd\omega\, g(\omega,\bfv)
&\text{for NH,}\\
&\\
\displaystyle\int\!\frac{\rmd\bfv}{4\pi}
\int_{-\infty}^0\rmd\omega\, g(\omega,\bfv)
&\text{for IH,}
\end{cases}
\end{align}
where 
$n_\nue$ ($n_\nux$) is the total number
density of $\nue$ ($\nux$), and NH and IH stand for the normal and
inverted neutrino mass hierarchies, respectively. Note that, according
to Eq.~\eqref{eq:w}, the neutrino has a positive vacuum oscillation
frequency $\omega$ if $\delta m^2>0$ and negative $\omega$ if  $\delta m^2<0$.
In the rest of the paper we will sometimes simply use the words ``spectrum'' and
``distribution'' to 
refer to the ELN energy or angular distributions.

We will examine the situation where neutrinos are almost in pure weak
interaction states so that the (reduced, traceless) neutrino density
matrix (in the weak interaction basis) is of the form
\begin{align}
\varrho_{\omega,\bfv} \approx
\begin{bmatrix}
1 & \epsilon_{\omega,\bfv} \\ \epsilon_{\omega,\bfv}^* & -1
\end{bmatrix},
\end{align}
where $|\epsilon|\ll1$. We will look for the criteria for the onset of
collective neutrino oscillations by performing 
the linear flavor-stability analysis in
which the terms of order $\epsilon^2$ or higher are ignored
\cite{Banerjee:2011fj}. 

In the absence of collisions neutrinos obey the equation of motion (EoM)
\cite{Sigl:1992fn} 
\begin{align}
(\partial_t + \bfv\cdot\bnabla)\varrho_{\omega,\bfv}
&= -\rmi[\sfH_0 + \sfH_{\nu\nu},\, \varrho_{\omega,\bfv}],
\label{eq:eom}
\end{align}
where $\sfH_0$ is the conventional neutrino oscillation Hamiltonian in
the absence of ambient neutrinos, and $\sfH_{\nu\nu}$ is the potential
due to neutrino-neutrino forward scattering.
We assume  a small neutrino mixing angle so that 
\begin{align}
\sfH_0 \approx (-\omega+\lambda)\frac{\sigma_3}{2},
\end{align}
where $\lambda=\sqrt2\Gf n_e$ is (a measure of the strength of) the
matter potential,
and $\sigma_3=\text{diag}[1,-1]$ is the third Pauli matrix.
For the situation that we are interested in,
\begin{align}
\sfH_{\nu\nu} =
\frac{\mu}{2} \int\!\frac{\rmd\bfv'}{4\pi}\int_{-\infty}^\infty\rmd\omega'\,
(1-\bfv\cdot\bfv') g' \varrho',
\end{align}
where $\mu=\sqrt2\Gf n_\nue$ is (a measure of the strength of) the neutrino
potential. For simplicity we sometimes suppress the arguments or
subscripts of certain physical quantities, and
the primed symbols such as $g'=g(\omega',\bfv')$ are understood as the
corresponding quantities that depend on
the integration variables $\omega'$ and $\bfv'$. As usual we have
ignored the trace terms which do not affect neutrino oscillations.

\section{Flavor instabilities in homogeneous and isotropic neutrino
  gases%
\label{sec:homo}}

It will prove helpful to first consider the relation between spectrum
crossing and the flavor instabilities in a
homogeneous and isotropic neutrino gas.
The importance of spectrum crossing was first pointed 
out by Dasgupta et al in a report of multiple spectral splits resulted from
collective neutrino oscillations \cite{Dasgupta:2009mg}. 
An explicit proof of this relation was later given by Banerjee et al
in an analysis on the flavor instabilities of dense neutrino gases
\cite{Banerjee:2011fj}.   
As a useful example we will inspect the flavor instabilities in the
bipolar model from the perspective of spectrum crossing.
For completeness we will also include a brief proof of this relation for a
neutrino gas with a continuous energy spectrum as that in
Ref.~\cite{Banerjee:2011fj} which
will be used later in the paper.

\subsection{Bipolar model%
\label{sec:bipolar}}

For a homogeneous and isotropic neutrino
medium one has
\begin{align}
\varrho_{\omega,\bfv}(t,\bfx) &= \varrho_\omega(t), &
g(\omega,\bfv) &= g(\omega)
\end{align}
and 
\begin{align}
\rmi\dot\varrho=\left[(-\omega+\lambda) \frac{\sigma_3}{2}
+ \frac{\mu}{2} \int_{-\infty}^\infty g'
  \varrho'\,\rmd\omega'
,\,
\varrho\right].
\label{eq:eom-homo}
\end{align}
The bipolar model 
describes a homogeneous and isotropic neutrino gas initially
consisting of mono-energetic $\nu_e$ and $\bar\nu_e$ (see, e.g.,
Refs~\cite{Kostelecky:1994dt,Hannestad:2006nj,Duan:2007mv}). 
It has a discrete energy spectrum
\begin{align}
g(\omega) = \sum_{i=1,2} g_i \delta(\omega-\omega_i),
\label{eq:g-discrete}
\end{align}
where $g_1=1$ and $\omega_1=\delta m^2/2 E_1$ for the neutrino with
energy $E_1$, and
$g_2=-n_{\bar\nu_e}/n_{\nu_e}$ and $\omega_2=-\omega_1$ for
the antineutrino. 
Keeping only the linear terms in the EoM one has
\begin{align}
\rmi\begin{bmatrix}
\dot\epsilon_1 \\ \dot\epsilon_2
\end{bmatrix}
=\begin{bmatrix}
-\omega_1 + \lambda + \mu g_2 & -\mu g_2 \\
-\mu g_1 & -\omega_2 + \lambda + \mu g_1
\end{bmatrix}  \begin{bmatrix}
\epsilon_1 \\ \epsilon_2
\end{bmatrix}.
\label{eq:eom-lin-bipolar}
\end{align}
In the linear regime collective neutrino oscillations are represented
by the normal modes
\begin{align}
\begin{bmatrix}
\epsilon_1(t) \\ \epsilon_2(t)
\end{bmatrix}
=\begin{bmatrix}
Q_1 \\ Q_2
\end{bmatrix} \exp(+\rmi\Omega t),
\end{align}
where $Q_i$ ($i=1,2$) are two complex constants, and
$\Omega$ is the collective oscillation frequency which can be
solved from the characteristic equation
\begin{align}
(\Omega -\omega_1 + \lambda + \mu g_2)(\Omega -\omega_2 +\lambda + \mu g_1)
  - \mu^2 g_1 g_2   = 0.
\label{eq:Omega-bipolar}
\end{align} 

The bipolar model becomes flavor unstable if there exists a solution with
\begin{align}
\kappa = -\text{Im}(\Omega) > 0.
\end{align}
From Eq.~\eqref{eq:Omega-bipolar} it is straightforward to show that a
flavor instability can exist only if
\begin{subequations}
\label{eq:cond-bipolar}
\begin{gather}
[(\omega_1-\omega_2) + (g_1-g_2)\mu]^2 
< -4 g_1 g_2 \mu^2,
\label{eq:cond-bipolar1}
\\
\intertext{or, equivalently,}
(g_1+g_2)^2\mu^2 + (\omega_1-\omega_2)^2 < 
- 2(\omega_1-\omega_2)(g_1-g_2)\mu. 
\label{eq:cond-bipolar2}
\end{gather}
\end{subequations}

\begin{figure}[ht]
  \begin{center}
    $\begin{array}{@{}c@{\hspace{0.3in}}c@{}}
      \includegraphics*[scale=0.55]{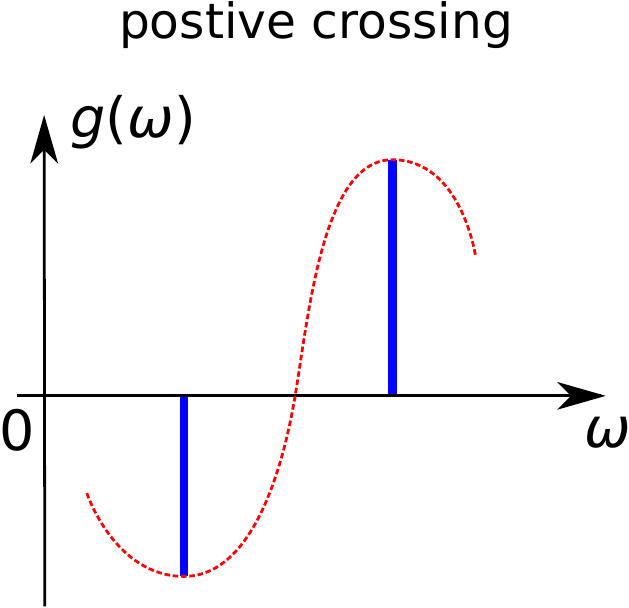} 
       & \includegraphics*[scale=0.55]{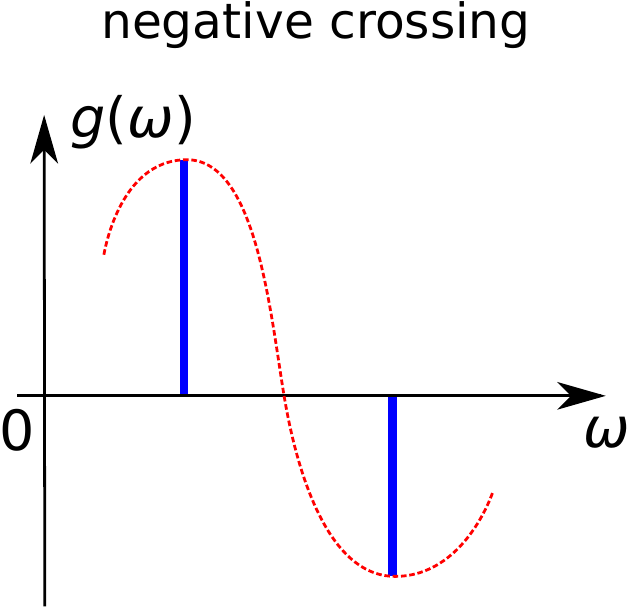}
      \end{array}$
  \end{center}
  \caption{Illustrative discrete (thick solid lines) and continuous
    (dashed curves) electron lepton number (ELN) spectra $g(\omega)$ with
    positive and negative 
    crossings, respectively, where $\omega$ is the oscillation frequency
    of the neutrino. For the discrete ELN spectrum, the lengths
    of the thick solid lines represent the coefficients in front of the
    delta functions in Eq.~\eqref{eq:g-discrete}.
    Collective oscillations can occur in a dense
    neutrino medium only if its initial ELN spectrum possesses
    one or multiple negative crossings.}
\label{fig:crossing}
\end{figure}

A few remarks are in order.
\begin{itemize}

\item Because inequalities~\eqref{eq:cond-bipolar} do not depend on
  $\lambda$, the presence of a uniform matter density does not affect flavor
  instability. In fact, it is obvious from
  Eq.~\eqref{eq:eom-lin-bipolar} that the presence of the ambient
  matter only shifts the collective oscillation frequency $\Omega$
  by $-\lambda$ \cite{Duan:2005cp}.

\item Eq.~\eqref{eq:eom-lin-bipolar} implies that,
  if the oscillation frequencies of both
  neutrino modes are shifted by a common value $\Delta\Omega$, then the
  collective oscillation frequency $\Omega$ is also shifted by
  $\Delta\Omega$ which
  again does not affect the flavor instability \cite{Duan:2005cp}.

\item Eq.~\eqref{eq:cond-bipolar1} implies that a neutrino flavor
  instability exists only if the ELNs of the two 
  neutrino modes, $g_1$ and $g_2$, have opposite signs, i.e.,
  $g_1g_2<0$.
  (See Fig.~\ref{fig:crossing}.)

\item Eq.~\eqref{eq:cond-bipolar2} gives another necessary condition
  of flavor instability,
  $(\omega_1-\omega_2)(g_1-g_2)<0$. For the bipolar system that we
  considers, this is possible only if $\omega_1<0$ which requires
  $\delta m^2<0$, i.e., the neutrino mass hierarchy is inverted.%
\footnote{In Ref.~\cite{Dasgupta:2009mg} and some other references,
  $\omega$ is defined to be
  always positive for the neutrino. In that convention, an energy spectrum with
  a negative (positive) crossing can be unstable if the neutrino has a
  normal (inverted) neutrino mass hierarchy.
  In our convention, however, $g(\omega)\rightarrow
  g(-\omega)$ as the neutrino mass hierarchy is flipped, and
  a negative crossing  in the energy
  spectrum becomes positive and vice versa.   The flavor
  stabilities will change accordingly. Either way, a neutrino system
  with a single crossed energy spectrum can be unstable with one neutrino
  mass hierarchy but stable with the other.
}

\end{itemize}

If the neutrino has the inverted mass hierarchy,
it is straightforward to show from Eq.~\eqref{eq:cond-bipolar} that
the bipolar model
is unstable for $\mu\in(\mu_-, \mu_+)$, 
where the lower and upper boundaries of the instability region are
\begin{align}
\mu_\pm = \frac{|\omega_1-\omega_2|}{(\sqrt{|g_1|}\mp\sqrt{|g_2|})^2},
\label{eq:mu-lim}
\end{align}
respectively. Unless $|g_1|$ and $|g_2|$ are almost the same, $\mu_+$
and $\mu_-$ 
are of the same  magnitude as the spread of the oscillation
frequencies $\Delta\omega=|\omega_1-\omega_2|$. By the dimension argument
one expects the maximum value of $\kappa$ to be of the same
magnitude as $\Delta\omega$, which in turn is usually of the same order as
$|\omega_1|$. Therefore, like the MSW mechanism, the flavor conversion
in the bipolar model, and also in
homogeneous and isotropic neutrino media in general,
usually occurs at neutrino densities $\lesssim \Gf^{-1}\omega$ and on 
time/distance scales of order $\omega^{-1}$ with $\omega$ being a characteristic
vacuum oscillation frequency of the neutrino gas.

\subsection{Neutrino gas of a continuous spectrum%
\label{sec:homo-cont-spec}
\footnote{The proof for the necessity of a crossed spectrum for
  neutrino flavor instability was first given by Banerjee et al in
  Ref.~\cite{Banerjee:2011fj}. The content of this section is included
for completeness.}  
}

The linearized EoM of a homogeneous and isotropic neutrino gas with a
continuous energy spectrum is
\begin{align}
\rmi\dot\epsilon =
(-\omega+\bar\lambda)\epsilon
-\mu\int_{-\infty}^\infty g'\epsilon'\,\rmd\omega',
\label{eq:eom-homo-lin}
\end{align}
where $\bar\lambda=\lambda+\Phi_0$ with
\begin{align}
\Phi_0 = \mu\int_{-\infty}^\infty g(\omega)\,\rmd\omega
= \sqrt2\Gf[(n_{\nue}-n_{\nux}) -
  (n_{\nueb}-n_{\nuxb})].
\end{align}
Assuming solution
$\epsilon_\omega(t) = Q_\omega e^{\rmi\Omega t}$ to
Eq.~\eqref{eq:eom-homo-lin} one obtains
\begin{align}
(\Omega+\bar\lambda - \omega) Q = 
\mu \int_{-\infty}^\infty g' Q'\,\rmd\omega',
\label{eq:Q}
\end{align}
which implies 
\begin{align}
Q_\omega\propto \frac{1}{\Omega+\bar\lambda-\omega}.
\label{eq:Q1}
\end{align}
Again one sees that, $\bar\lambda$, a combined potential due to the
matter and neutrino background, has the effect 
of shifting the collective oscillation frequency $\Omega$ but does not
affect the flavor instability. For the purpose of analyzing flavor
instabilities we will assume $\bar\lambda=0$ in this section.
Substituting Eq.~\eqref{eq:Q1} back into Eq.~\eqref{eq:Q} one obtains
\begin{align}
\int_{-\infty}^\infty \frac{g(\omega)\,\rmd\omega}
{(\omega -\gamma)+\rmi\kappa}
=-\frac{1}{\mu},
\label{eq:homo-sta}
\end{align}
where $\gamma = \text{Re}(\Omega)$.

Suppose that there exists a flavor
instability for $\mu$ within the range of $(\mu_-,\mu_+)$. In the limit
$\mu\rightarrow \mu_- + 0^+$, $\kappa\rightarrow 0^+$, and
Eq.~\eqref{eq:homo-sta} becomes 
\begin{align}
\mathcal{P}\int_{-\infty}^\infty \frac{g(\omega)\,\rmd\omega}
{\omega - \wc} -\rmi\pi g(\wc) = -\frac{1}{\mu_-},
\label{eq:homo-cross}
\end{align}
where $\mathcal{P}$ signifies the Cauchy principal value of the integral,
and  $\wc=\Omega|_{\mu=\mu_-}$.%
\footnote{Here we have assumed that, if $g(\omega)$ is nonzero only within range
$(\omega_\text{min}, \omega_\text{max})$, then
$\omega_\text{min}<\wc< \omega_\text{max}$. This is a reasonable
assumption because $\mu_-$ is small and, therefore, collective
oscillation frequency $\Omega$ at $\mu=\mu_-$ is an ``average'' value
of the oscillation frequency $\omega$ which should be
within the range of $(\omega_\text{min}, \omega_\text{max})$.} 
Because $\mu$ is real, Eq.~\eqref{eq:homo-cross}
implies that the spectrum $g(\omega)$ must cross 0 at $\omega=\wc$.
In addition, unless there are other crossing
points nearby, the integral in Eq.~\eqref{eq:homo-cross} has the same
sign as the crossing of $g(\omega)$ at $\wc$. This in turn
implies that the spectrum crossing at $\wc$ must be negative because
we have defined 
$\mu$ to be positive.
This result can be considered as a generalization of that in the
bipolar model.   (See Fig.~\ref{fig:crossing}.)

\section{Fast flavor conversions in anisotropic neutrino media%
\label{sec:aniso}}

We will now turn to the relation between fast flavor conversions and
the crossings in the neutrino angular distribution. 
The first, albeit inconclusive, observation of this relation was
made by Dasgupta et al in a study of fast neutrino
oscillations in a multi-bulb supernova model \cite{Dasgupta:2016dbv}.
Here we will consider stationary, anisotropic neutrino media
which consist of 
neutrino fluxes emitted constantly from planar surfaces. We
assume that the neutrino emission is uniform on the emission plane
and that the matter distribution is homogeneous in the whole space.
We will first use the four-beam neutrino model to demonstrate the
origin of fast flavor conversions by relating it to the bipolar model.
We will then give an explicit proof of the necessity of a crossed
neutrino angular distribution for fast oscillations to occur in an
outward flowing, axially symmetric neutrino flux.

\subsection{Four-beam model}

\begin{figure}[ht]
  \begin{center}
    \includegraphics*[scale=0.6]{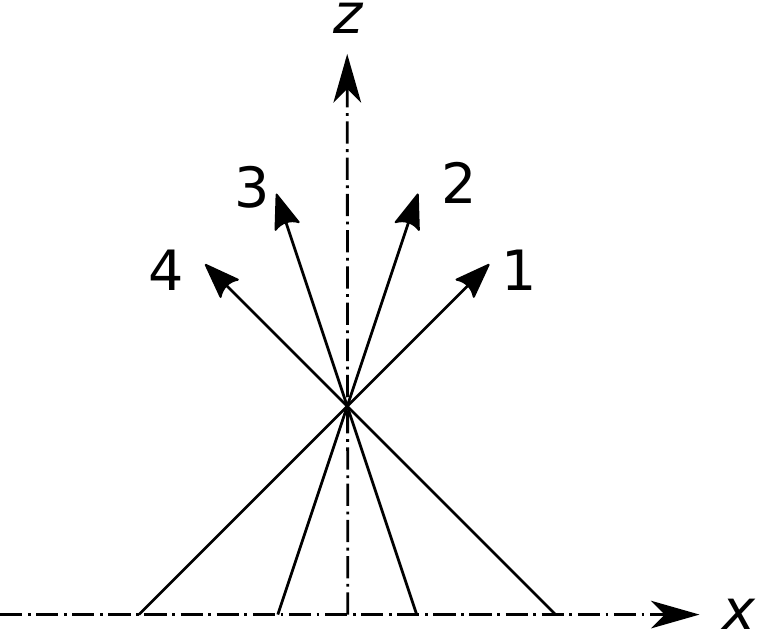} 
  \end{center}
  \caption{A schematic diagram of the four-beam model in which the
    neutrinos of four momentum modes are emitted constantly and
    homogeneously from the $x$-$y$ plane in the directions that are
    parallel to the $x$-$z$ plane.}
\label{fig:4beam}
\end{figure}

The four-beam neutrino model has the ELN distribution
\begin{align}
g(\bfv,\omega) = 4\pi\sum_{i=1}^4 g_i \delta(\bfv-\bfv_i)\delta(\omega-\omega_i),
\end{align}
where $\bfv_i$ and $\omega_i$ ($i=1,2,3,4$) are the velocities and
oscillation frequencies of the four neutrino modes of the system. 
(See Fig.~\ref{fig:4beam}).  We
will assume the reflection symmetry between
neutrino modes $(1, 2)$  and $(4, 3)$ 
in the four-beam model except for small differences among $\varrho_i$, i.e.,
$ (\epsilon_1, \epsilon_2)\neq(\epsilon_4, \epsilon_3)$. We define
\begin{align}
\epsilon_{1\pm}(z) &= \frac{\epsilon_1(z)\pm\epsilon_4(z)}{2}, &
\epsilon_{2\pm}(z) &= \frac{\epsilon_2(z)\pm\epsilon_3(z)}{2},
\end{align}
where the $z$ direction is perpendicular to the neutrino emission
plane.
Because of the reflection symmetry in the EoM, the 
symmetric modes ($\epsilon_{i+}$) and the anti-symmetric modes
($\epsilon_{i-}$) evolve independently  in the linear regime.

\subsubsection{Symmetric mode}

\begin{figure*}[htb]
  \begin{center}
    $\begin{array}{@{}c@{\hspace{0.05in}}c@{}}
      \includegraphics*[scale=0.7]{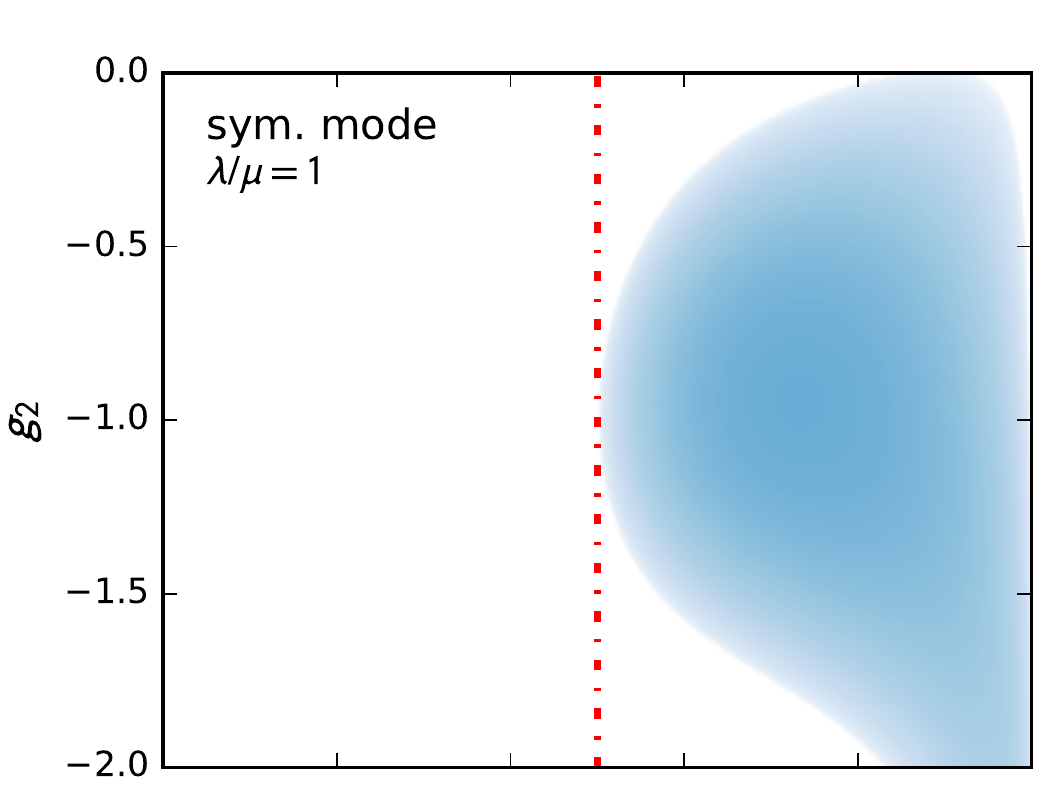} &
      \includegraphics*[scale=0.7]{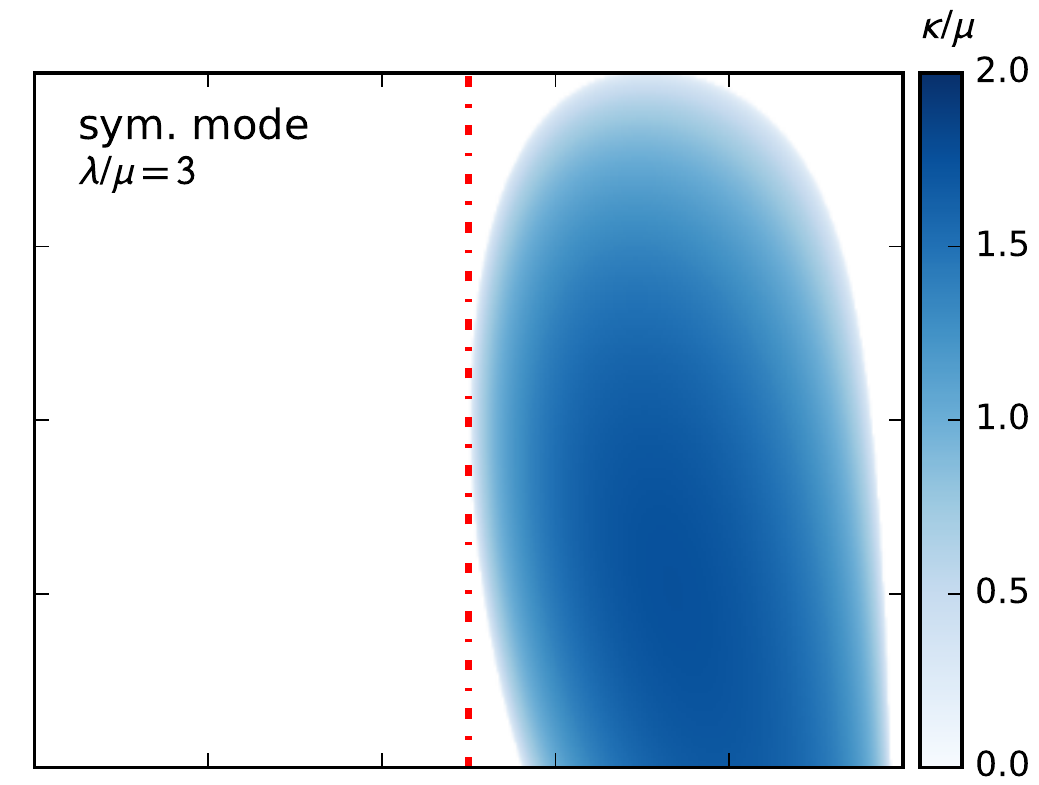} \\
      \includegraphics*[scale=0.7]{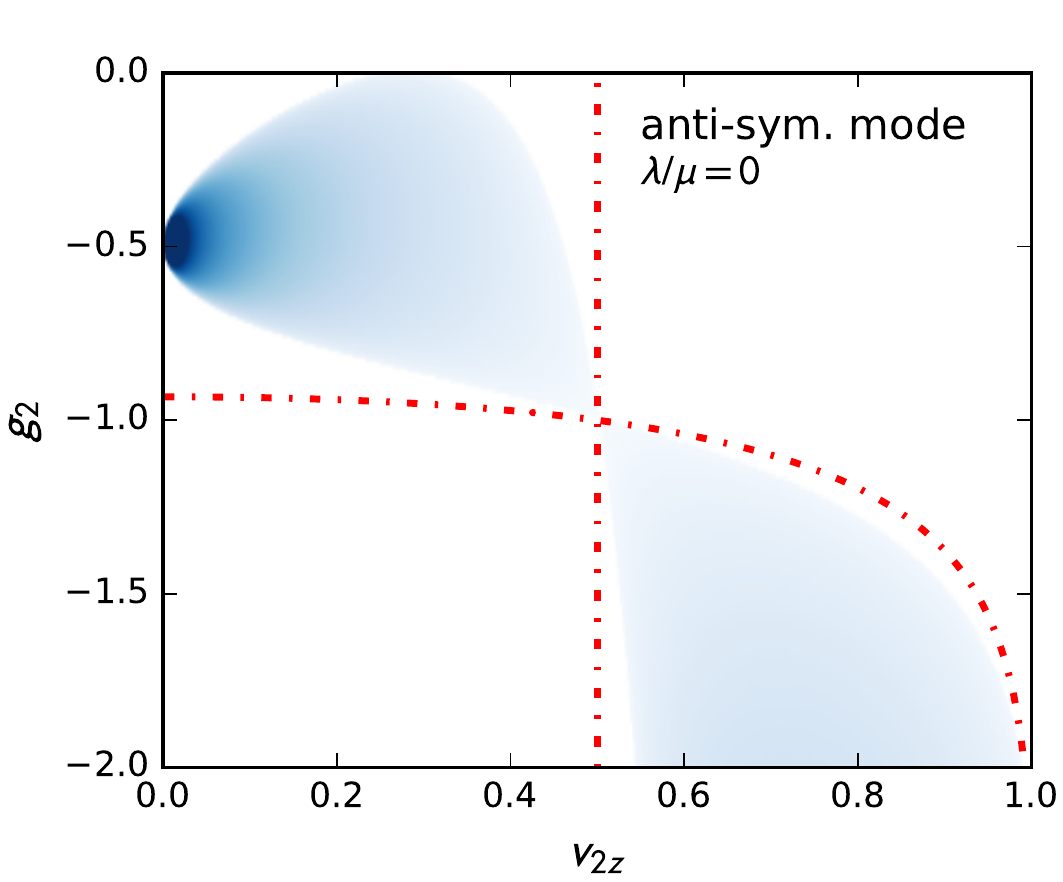} &
      \includegraphics*[scale=0.7]{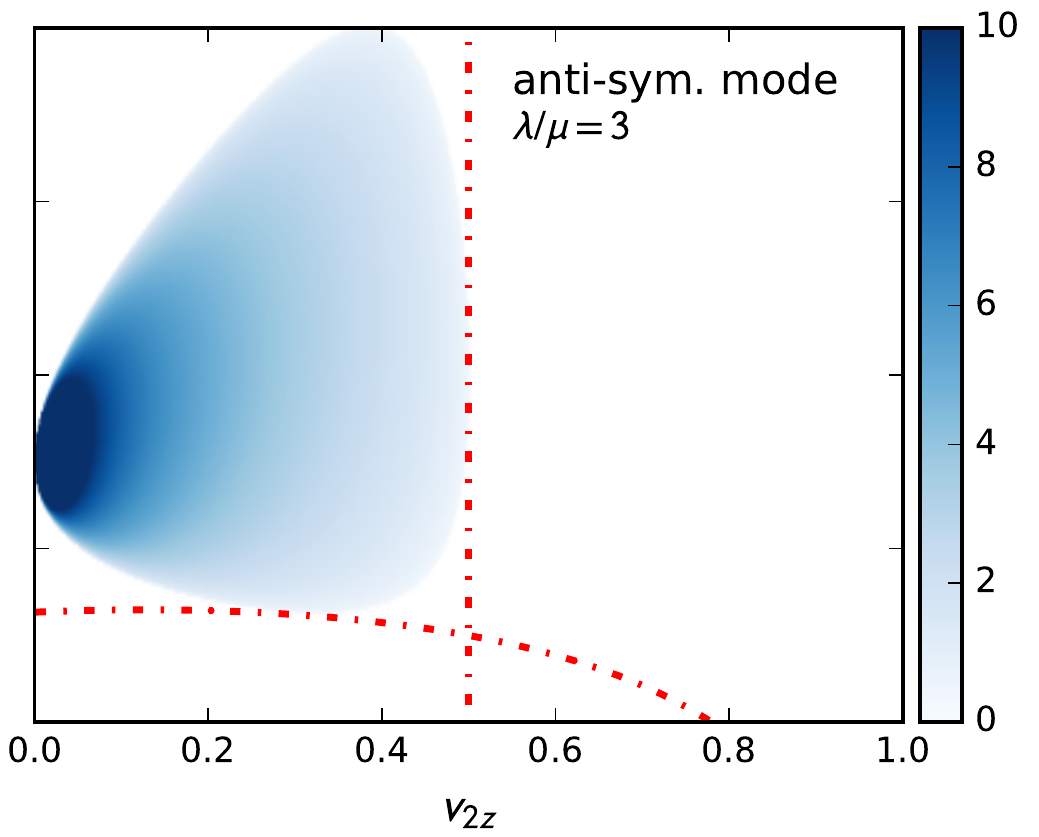}
      \end{array}$
  \end{center}
  \caption{Flavor instabilities in the four-beam neutrino model in
    Fig.~\ref{fig:4beam} 
    for the neutrino oscillation modes which preserve the reflection
    symmetry (top 
    panels) and break this symmetry (bottom panels), respectively. The
    color density indicates
    the exponential growth rate $\kappa$ of the flavor instability of the
    neutrino medium as a function of the
    electron lepton numbers (ELNs) $g_2=g_3$ and the $z$-component
    velocities $v_{2z}=v_{3z}$ of two of the neutrino modes.
    The other two modes have fixed values of $g_1=g_4=1$ and
    $v_{1z}=v_{4z}=1/2$. The dot-dashed curves mark the boundaries between
    the scenarios where the effective ELN spectra 
    have negative and 
    positive crossings, respectively. Both $\kappa$ and the strength
    of the matter potential 
    $\lambda$ (as indicated) are measured in the strength of the
    neutrino potential $\mu$.}
 \label{fig:kappa-line}
\end{figure*}

The symmetric mode of the four-beam model obeys the EoM
\begin{align}
\rmi\td{z}\begin{bmatrix}
\epsilon_{1+} \\ \epsilon_{2+}
\end{bmatrix}
=\begin{bmatrix}
-\tomega_{1+} + \tmu_+ \tg_{2+} & -\tmu_+ \tg_{2+} \\
-\tmu_+ \tg_{1+} & -\tomega_{2+} + \tmu_+ \tg_{1+}
\end{bmatrix}  \begin{bmatrix}
\epsilon_{1+} \\ \epsilon_{2+}
\end{bmatrix}
\label{eq:eom-4b}
\end{align}
in the linear regime, where 
\begin{subequations}
\label{eq:par+}
\begin{align}
\tomega_{1+} &= \frac{\omega_1-\lambda}{v_{1z}}, &
\tg_{1+} &= \frac{g_1}{v_{2z}}, \\
\tomega_{2+} &= \frac{\omega_2-\lambda}{v_{2z}}, &
\tg_{2+} &= \frac{g_2}{v_{1z}}, \\
\tmu_+ &= \mu_{12}+\mu_{13} & &
\end{align}
\end{subequations}
with $v_z$ being the $z$ component of $\bfv$, and
$\mu_{ij}=(1-\bfv_i\cdot\bfv_j)\mu$. 
Here we have rewritten the EoM of the symmetric mode in a form similar
to that of the bipolar model, Eq.~\eqref{eq:eom-lin-bipolar}, so that
the results of the bipolar model can be applied here.
Eq.~\eqref{eq:par+} shows that the presence of the ordinary matter
can contribute to the spread of effective oscillation frequencies
$\Delta\tomega=|\tomega_{1+}-\tomega_{2+}|$, which is generally true for a 
system with an anisotropic neutrino emission. This
multi-angle matter effect is responsible for the suppression of neutrino
oscillations in the 
accretion phase of a core-collapse supernova if all neutrino species
are emitted from the same neutrino sphere and with the same angular
distribution \cite{EstebanPretel:2008ni}. 

Here we want to emphasize that the same multi-angle matter effect can
also induce fast flavor conversions.
To see this we take the limit 
$\delta m^2\rightarrow 0$ so that
\begin{align}
\tomega_{1+}&\rightarrow -\frac{\lambda}{v_{1z}}, &
\tomega_{2+}&\rightarrow -\frac{\lambda}{v_{2z}}.
\label{eq:omega+}
\end{align}
The discussion in Sec.~\ref{sec:bipolar} implies that there can exist
flavor instability even in this limit. Assuming $v_{1z} < v_{2z}$, the
requirement of a negatively crossed effective spectrum
$\tg_+(\tomega_+)$ means that 
$g_1>0>g_2$, e.g., beams $(1, 4)$ are emitted in the $\nue$ state and
beams $(2, 3)$ in the $\nueb$ state. In other words, the existence of
a flavor instability in the symmetric mode requires that there exist a
negative crossing in the angular distribution 
\begin{align}
G(v_z) = \int_{-\infty}^\infty g(v_z,\omega)\,\rmd\omega.
\label{eq:G}
\end{align}
When this is the case the neutrino medium becomes unstable against the flavor
conversion when 
$\mu$ is comparable to $\lambda$. [See Eqs.~\eqref{eq:mu-lim} and
\eqref{eq:omega+}.] The
exponential growth rate $\kappa$ of the flavor instability is also of the
same order of $\lambda$ which can be much larger than vacuum
oscillation frequencies (taken to be zero in this limit).

As an example we consider the case where beams $(1, 4)$ are
emitted in the $\nu_e$ state with $v_{1z}=1/2$ and $g_1=1$ and beams $(2, 3)$ in
the $\bar\nu_e$ state with various values of $v_{2z}$ and $g_2$. In
the two top panels of Fig.~\ref{fig:kappa-line} we plot
the exponential growth rate $\kappa$ of the flavor instability as
a function of $v_{2z}$ and $g_2$ with fixed ratios $\lambda/\mu=1$ and $3$,
respectively. We only plotted the results with $g_2<0$ where
there exits a crossing in the effective spectrum
$\tg_+(\tomega_+)$. From the figure one sees that the
symmetric fast flavor conversions
are possible  in the four-beam model only if the
$\bar\nu_e$ beams are emitted in a more 
forward direction than the $\nu_e$ beams so that $\tg_+(\tomega_+)$ has a
negative crossing.

\subsubsection{Anti-symmetric mode}

The EoM for the anti-symmetric mode is the same as Eq.~\eqref{eq:eom-4b}
except that the ``plus'' quantities are replaced by their ``minus''
counterparts:
\begin{subequations}
\label{eq:par-}
\begin{align}
\tomega_{1-} &= \frac{\omega_1-\lambda-2\mu_{14}g_1-2\mu_{13}g_2}{v_{1z}}, &
\tg_{1-} &= -\frac{g_1}{v_{2z}}, \\
\tomega_{2-} &= \frac{\omega_2-\lambda-2\mu_{13}g_1-2\mu_{23}g_2}{v_{2z}}, &
\tg_{2-} &= -\frac{g_2}{v_{1z}}, \\
\tmu_- &= -\mu_{12}+\mu_{13}. & &
\end{align}
\end{subequations}
Without losing generality we will assume the configuration with
$\bfv_1\cdot\bfv_2>\bfv_1\cdot\bfv_3$ so that $\tmu_->0$.
The comparison between Eq.~\eqref{eq:par-} and Eq.~\eqref{eq:par+} reveals
two important features of the anti-symmetric collective mode in the
four-beam model:
\begin{itemize}
\item Both the ordinary matter and the background neutrino medium
contribute to the spread of effective
oscillation frequencies $\Delta\tomega_-=|\tomega_{1-}-\tomega_{2-}|$.
This multi-angle effect due to the neutrino medium itself can suppress
collective neutrino oscillations in the supernova model in which all
neutrino species have the same angular distribution \cite{Duan:2010bf}.

\item The effective spectrum $\tg_-(\tomega_-)$ has a sign opposite
  to that of the actual ELN distribution. This implies that a flavor
  instability may exist when there is a positive crossing in the 
  angular distribution $G(v_z)$. 
\end{itemize}

As in the case of the symmetric mode, there can exist a flavor instability
for the anti-symmetric mode in the limit $\delta
m^2\rightarrow 0$. A crossing in the effective ELN
spectrum $\tg_-(\tomega_-)$  (i.e.\ $\tg_{1-}\tg_{2-}<0$) also implies
a crossing in the actual angular distribution $G(v_z)$ ($g_1g_2<0$).
Unlike the symmetric mode, the anti-symmetric mode
can have fast flavor conversions even when 
$\lambda=0$. In addition,
the multi-angle neutrino effect can change the
the order of $\tomega_{i-}$, i.e.\ whether $\tomega_{1-}<\tomega_{2-}$
or $\tomega_{1-}>\tomega_{2-}$. As a result, the regimes where the
effective spectrum $\tg_-(\tomega_-)$ has a negative crossing are
not solely determined by 
$v_{2z}$ as in the case of the symmetric mode.
In the lower panels of Fig.~\ref{fig:kappa-line} we plot
the exponential growth rate $\kappa$ of this instability as
a function of $v_{2z}$ and $g_2$ with fixed ratios $\lambda/\mu=0$ and $3$,
respectively. 

\subsection{Outward flowing, axially symmetric neutrino flux%
\label{sec:outflux}}

We next consider an outward flowing,
nearly axially symmetric neutrino flux with 
\begin{align}
g(\bfv,\omega) = g(v_z,\omega)
\end{align} 
and small perturbations to the initial states of the neutrinos which
can depend the azimuthal angle $\varphi$ of $\bfv$.
We will again consider flavor instabilities in the limit $\delta
m^2\rightarrow 0$.
The linearized EoM of this model is
\begin{align}
\rmi v_z \tdf{z}{\epsilon} 
= (\lambda+\Phi_0 - v_z \Phi_z) \epsilon
-\mu\int(1-\bfv\cdot\bfv') G' \epsilon' \,\frac{\rmd\bfv'}{4\pi},
\label{eq:eom-plane}
\end{align}
where 
\begin{align}
\Phi_0&= \mu\int G'\,\frac{\rmd\bfv'}{4\pi}, &
\Phi_z &= \mu\int v'_z G'\,\frac{\rmd\bfv'}{4\pi}.
\end{align}
Following Ref.~\cite{Banerjee:2011fj} we assume a solution of the form
\begin{align}
\epsilon_{\bfv}(z) = \frac{Q_{\bfv}}{v_z} e^{\rmi \Omega z}
\end{align} 
to Eq.~\eqref{eq:eom-plane} and obtain
\begin{align}
\left(\Omega -\Phi_z + \frac{\bar\lambda}{v_z}\right)  Q &= 
\mu \int (1-\bfv\cdot\bfv') G'  \frac{Q'}{v'_z}\,\frac{\rmd\bfv'}{4\pi},
\label{eq:Q-plane}
\end{align}
where $\bar\lambda=\lambda+\Phi_0$.
Comparing Eqs.~\eqref{eq:Q} and \eqref{eq:Q-plane}
one sees that $\Phi_z$ shifts the collective
oscillation frequency $\Omega$ while $-\bar\lambda/v_z$ plays the role of
vacuum oscillation frequency $\omega$.
Eq.~\eqref{eq:Q-plane} implies 
\begin{align}
Q_{\bfv} = \frac{A + B v_z +  C \sqrt{u} \cos\varphi + D \sqrt{u} \sin\varphi}%
{\Omega-\Phi_z + \bar\lambda/v_z},
\label{eq:Qv}
\end{align}
where 
\begin{subequations}
\label{eq:ABCD}
\begin{align}
A &= \mu\int G' \frac{Q'}{v'_z}\, \frac{\rmd \bfv'}{4\pi}, \\
B &= -\mu\int G' Q' \,\frac{\rmd \bfv'}{4\pi},\\
C &= -\mu\int \sqrt{u'} \cos\varphi'\, 
G' \frac{Q'}{v'_z}\,\frac{\rmd \bfv'}{4\pi},\\
D &= -\mu\int \sqrt{u'}\sin\varphi'\, 
G' \frac{Q'}{v'_z}\,\frac{\rmd \bfv'}{4\pi},
\end{align}
\end{subequations}
and $u=1-v_z^2$.
Substituting Eq.~\eqref{eq:Qv} to Eq.~\eqref{eq:ABCD} we obtain
\begin{align}
\begin{bmatrix}
I[1]-1  & I[v_z] & 0 & 0 \\
-I[v_z] & -I[v_z^2]-1  & 0 & 0 \\
0 & 0 & -\frac{1}{2}I[u]-1  & 0 \\
0 & 0 & 0 & -\frac{1}{2}I[u] -1
\end{bmatrix}
\begin{bmatrix}
A \\ B \\ C \\ D
\end{bmatrix}=0,  
\label{eq:cha}
\end{align} 
where 
\begin{align}
I[f(v_z)] = \frac{\mu}{2} \int_0^1 \frac{G(v_z) f(v_z)}
{\Omega-\Phi_z +\bar\lambda/v_z}\,\frac{\rmd v_z}{v_z} .
\end{align}
Eq.~\eqref{eq:cha} has a solution only if
\begin{subequations}
\label{eq:I}
\begin{align}
(I[1] -1)(I[v_z^2]+1)&=I^2[v_z]
\label{eq:Ia}
\intertext{or}
I[u] &= -2,
\label{eq:Ib}
\end{align}
\end{subequations}
which corresponds to the collective modes that preserve and break the
axial symmetry, respectively.
\footnote{The derivation of the characteristic equation \eqref{eq:I} for this
model is similar to that  in 
Refs.~\cite{Chakraborty:2016lct,Dasgupta:2016dbv} and is included
here for completeness.}

Let us consider the symmetry-breaking solution first. We rewrite
Eq.~\eqref{eq:Ib} as
\begin{align}
\int_{\tomega_\text{min}}^{\tomega_\text{max}}
\frac{\tg(\tomega)\,\rmd\tomega}
{[\tomega-(\gamma-\Phi_z)]+\rmi\kappa} = \frac{1}{\mu},
\label{eq:MAA-sta}
\end{align}
where 
\begin{align}
\tomega &= -\frac{\bar\lambda}{v_z}, &
\tg(\tomega) &= v_z(1-v_z^2) \frac{ G(v_z)}{4\bar\lambda},
\end{align}
and $(\tomega_\text{min},\tomega_\text{max})=(-\bar\lambda,\infty)$
if $\bar\lambda<0$ and $(-\infty,-\bar\lambda)$ if $\bar\lambda>0$.
Using the result of the homogeneous, isotropic medium (in
Sec.~\ref{sec:homo-cont-spec}), we determine that fast flavor conversions
that break the axial symmetry can occur in this medium
only if there exists a positive crossing in 
the  angular distribution $ G(v_z)$ within $[0,1]$.

One can also perform a similar analysis to Eq.~\eqref{eq:Ia} and show
that fast flavor 
conversions that preserve the axial symmetry can take place
if there exists a crossing in $ G(v_z)$, although the
sign of the crossing is not constrained. (See Appendix \ref{sec:appendix}
for details.)

\section{Multi-bulb supernova model%
\label{sec:2bulb}}

Now we consider the simple supernova model proposed by Sawyer in
Ref.~\cite{Sawyer:2015dsa} in which $\nue$, $\nueb$, $\nux$ are
emitted constantly and
half-isotropically from three spherically symmetric neutrino spheres
or ``bulbs'' of radii $R_\nue$,
$R_\nueb$ and $R_\nux$, respectively, and $\nuxb$ has the identical
emission characteristics as $\nux$.
This multi-bulb model is a
generalization of the neutrino (single-)bulb model in
Ref.~\cite{Duan:2006an} where  $R_\nue=R_\nueb=R_\nux$. 
Because the typical length scale of the fast flavor conversion is
much smaller than $R_\nu$ ($\nu=\nue,\nueb,\nux$), we can ignore
the spherical nature of the model and apply the results of
Sec.~\ref{sec:outflux}.

In the multi-bulb model the 
number flux of the neutrino species 
at radius $r$ takes the form 
\begin{align}
j_\nu(r,\vartheta) = \frac{n_{\nu,0}}{2\pi} \Theta(R_\nu - r\sin\vartheta),
\end{align}
where $\vartheta$ is the polar angle of the flux with respect to the
radial direction, and $n_{\nu,0}$ is the number density of the
neutrino species $\nu$ on the
corresponding neutrino sphere. The total number luminosity of neutrino
species $\nu$ is 
\begin{align}
\dN_\nu = 4\pi r^2 \int_0^{2\pi}\rmd\varphi\int_0^1\rmd(\cos\vartheta)\,
  j_\nu(r,\vartheta)\cos\vartheta
= 2\pi R^2 n_{\nu,0}.
\end{align}

For $R_\nue>R_\nueb>R_\nux$ and $\dN_\nux:\dN_\nueb:\dN_\nue=0.62:0.77:1$
Sawyer found that there  could
exist a flavor instability even on the $\nue$ sphere if
\begin{align}
\frac{R_\nueb}{R_\nue}>0.44 + 0.55
\left(\frac{R_\nux}{R_\nue}\right).
\label{eq:sawyer-crit}
\end{align} 
For $R_\nue=R_\nueb/0.93=R_\nux/0.8 = 15~\text{km}$ and
$n_{\nueb,0}=10^{32}~\text{cm}^{-3}$
he calculated the instability growth rate to be $\kappa=0.32\text{
  m}^{-1}$ on the surface of the $\nu_e$ sphere. However, 
since $j_\nux=j_\nuxb$ in the multi-bulb model and, therefore,
the contribution
of $\nux$ and $\nuxb$ to the angular distribution $G(v_z)$ cancel
each other, the criterion for
``pure'' fast neutrino flavor conversions (with $\delta m^2\rightarrow
0)$ should not depend on $R_\nux$ or $n_{\nux,0}$, which is
contrary to the condition set forth in Eq.~\eqref{eq:sawyer-crit}.
(For this
reason the multi-bulb model was called the 
two-bulb model by Chakraborty et al in Ref.~\cite{Chakraborty:2016lct}.)
Further, it was observed by Dasgupta et al \cite{Dasgupta:2016dbv}
that, in order for fast  
flavor conversions to take place in the multi-bulb model, there should
exist at least one crossing in the angular distribution. This requirement
has been explicitly shown in Sec.~\ref{sec:outflux}, and implies that
\begin{align}
(n_{\nue,0}-n_{\nueb,0})(R_{\nue}-R_{\nueb}) < 0.
\label{eq:bulb-cross}
\end{align} 
The example given above by Sawyer in
Ref.~\cite{Sawyer:2015dsa} does not satisfy this criterion.
He also did not include the matter
effect which can significantly affect the fast neutrino flavor conversion. 
We have performed a linear flavor stability analysis to the
multi-bulb model, and we did not find any flavor instability
with the parameters given in
Ref.~\cite{Sawyer:2015dsa}.

In Ref.~\cite{Chakraborty:2016lct} Chakraborty et al applied the linear flavor
instability analysis to the multi-bulb model but only in the
far regime where $r\gg R_\nu$. In this regime
\begin{align}
2\pi j_\nu(r,\vartheta) \rmd(\cos\vartheta)
\approx n_\nu(r,u) (-\rmd u),
\end{align}
where 
$u=\sin^2\vartheta|_{r=R_\star}$,
and $n_\nue$ and $n_\nueb$ are parameterized as
\begin{align}
n_\nu \propto \frac{1\pm a}{1\pm b}\times\left\{\begin{array}{ll}
1 & \text{ for } 0 < u < 1\pm b,\\
0 & \text{ otherwise}
\end{array}\right.
\end{align} 
with the upper and lower signs applying to $\nu_e$ and $\nueb$,
respectively. In the above 
parameterization, the asymmetry parameter $a$ and the width parameter $b$
are both within the range of $(-1,1)$ and determine
$\dot{N}_\nue/\dot{N}_\nueb$ and $R_\nue/R_\nueb$, respectively, and $R_\star$
is the radius at which the above parameterization is valid. In this
parameterization, the angular distribution has a crossing only if 
\begin{align}
(a-b) b < 0.
\end{align}
Indeed, Figs.~5 and 6 of
Ref.~\cite{Chakraborty:2016lct} show that the flavor
instabilities exist only in the parameter space where the above
criterion is satisfied. Those figures also demonstrate that a large
matter density (i.e.\ $\lambda\gtrsim\mu$) can dramatically affect the
neutrino flavor instabilities in the multi-bulb model.

\begin{figure}[ht]
  \begin{center}
    \includegraphics*[width=\columnwidth]{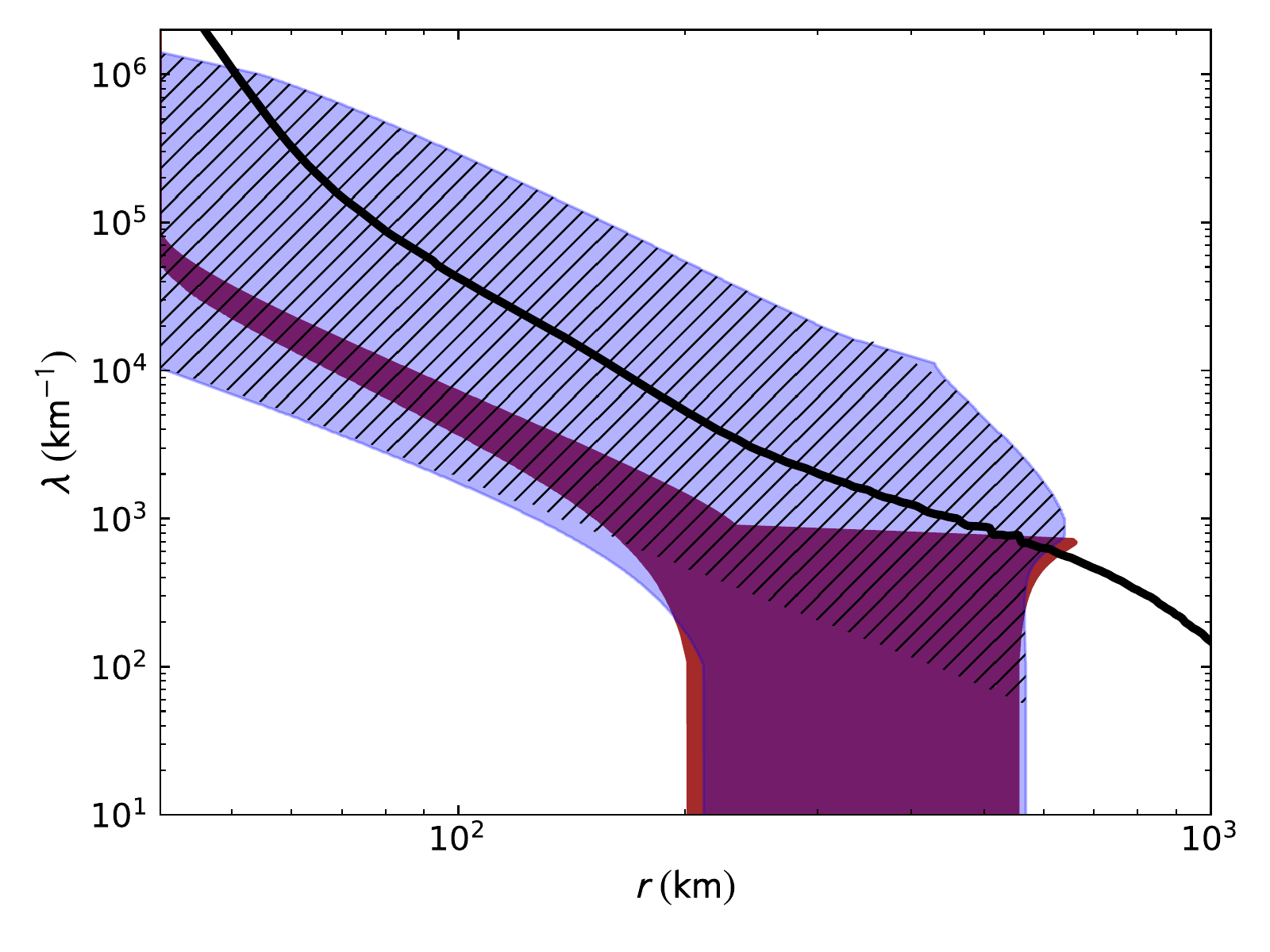}
  \end{center}
  \caption{The flavor instability region of the neutrino gas
    (in the light shadow) with various matter densities
    (proportional to the matter potential $\lambda$) 
    in a one-dimensional, electron-capture supernova
    model at $t\approx200$ ms after the core bounce
    \cite{Huedepohl:2009wh}. The thick solid 
    curve is the actual matter profile as a function of radius
    $r$. The hatched region is where the ``pure''
    fast flavor conversions (with the neutrino mass-squared splitting $\delta
    m^2\rightarrow 0$) become possible,
    and the dark shadowed region is where collective neutrino
    oscillations can occur if the neutrino spheres of
    $\nux$ and $\nueb$ coincide with that of $\nue$.}
 \label{fig:kappa-SN}
\end{figure}

As another example of the multi-bulb model, we calculated the neutrino
flavor instability region 
in a one-dimensional, electron-capture supernova model with an
$8.8M_\odot$  progenitor which was computed by the 
Garching group \cite{Huedepohl:2009wh}. At $t\approx200~\text{ms}$
after the core bounce this model has
$\dN_\nue=9.80\times10^{56}~\rms^{-1}$,
$\dN_\nueb=8.36\times10^{56}~\rms^{-1}$,
$\dN_\nux=6.92\times10^{56}~\rms^{-1}$, $R_\nue=40.2~\km$,
$R_\nueb=36.6~\km$ and $R_\nux=36.1~\km$, respectively. It is easy to
check that the neutrino gas in this supernova model has a crossed 
angular distribution.
Using the results in Sec.~\ref{sec:outflux} we plot the 
flavor unstable region in the supernova 
as the light shadowed area in Fig.~\ref{fig:kappa-SN} for
various matter densities.
To simplify the calculation, we have assumed the two-flavor
mixing with the inverted neutrino mass hierarchy and a single vacuum oscillation
frequency $\omega_0=0.68\,\text{km}^{-1}$ for all neutrino
species. This simplification should not have a dramatic impact on the
flavor unstable region in Fig.~\ref{fig:kappa-SN}
which is a log-log plot.
We depict the flavor unstable region of the pure fast modes
with $\delta m^2\rightarrow 0$ as the hatched area in the same figure.
As comparison we also plot the flavor unstable
region as dark shadowed region for the case where
the neutrino spheres of all other neutrino species coincide with that
of $\nue$. This corresponds to the single-bulb model where only
``slow'' neutrino 
oscillations can exist since there is no crossing in the angular
distribution. Finally we plot the actual matter profile as the thick
solid curve in the figure.

Fig.~\ref{fig:kappa-SN} clearly demonstrates the important influence of the
ambient matter on neutrino oscillations in a supernova. In the absence
of a dense matter background, collective neutrino oscillations are
self-suppressed near the neutrino spheres where the neutrino density
is large \cite{Duan:2010bf}. In the single-bulb model where all
neutrino species have an identical angular emission, collective
neutrino oscillations can take place but only in the parameter space
where the matter 
potential is finely matched with the neutrino potential which is not
the case in our example. In contrast, in the multi-bulb model which
has a crossed neutrino angular distribution, fast flavor
conversions can occur near the neutrino spheres for 
a broad range of matter densities spanning over two orders of
magnitude. Since the actual  
matter profile falls into the regime of the flavor instability in this
particular example, we
expect that the fast flavor conversion can develop there.

\section{Discussion and conclusions%
\label{sec:conclusions}}

We have used the four-beam neutrino model to explain the origin of
fast neutrino oscillations in a dense, anistropic neutrino gas even
in the limit that the 
neutrino mass splitting $\delta m^2\rightarrow 0$, although a finite
flavor mixing is necessary to initiate such a flavor conversion. 
It is the multi-angle effect of the neutrinos as
they travel through a dense matter and neutrino background that induce
the fast flavor conversion. This
flavor conversion can take place inside a very dense medium over very short
distance scales of order $(\Gf n_e)^{-1}$ or $(\Gf n_\nu)^{-1}$.

We have shown that the fast flavor conversion is generally prohibited
in a stationary, outward flowing, axially symmetric neutrino flux
unless there is a crossing in the neutrino angular
distribution. This provides an explicit proof for a similar conjecture
by Dasgupta et al in Ref.~\cite{Dasgupta:2016dbv} for the multi-bulb
supernova model. However, this conclusion does
\emph{not} apply to a medium where both inward and outward flowing
neutrino fluxes are present. This can be easily seen from the
four-beam model when  the directions of two of the neutrino
beams are flipped, e.g., \ $v_{2z}\rightarrow -v_{2z}$ and
$v_{3z}\rightarrow -v_{3z}$. This would result a sign change in
the effective spectrum 
\[ \tg_{1\pm}=\pm\frac{g_1}{v_{2z}}\rightarrow -\tg_{1\pm} \]
[see Eqs.~\eqref{eq:par+} and \eqref{eq:par-}].
Therefore, even if there is no crossing in the actual angular
distribution $G(v_z)$, there can still be a crossing in the effective
spectrum $\tg(\tomega)$ because of
this sign change in the part of the effective
spectrum for the inward neutrino fluxes. In other words, fast
flavor conversions can exist in a stationary neutrino gas with both inward and
outward neutrino fluxes, e.g.\ inside the $\nue$ sphere of a
core-collapse supernova, even if there is no crossing in the neutrino
angular distribution.

We note that the convective operator in the EoM
\eqref{eq:eom} becomes 
\[ \partial_t+\bfv\cdot\bnabla \longrightarrow
\rmi(-\Omega+\bfv\cdot\bfK) \]
for a collective oscillation wave of collective frequency $\Omega$
and wave vector $\bfK$ \cite{Duan:2008v1,Izaguirre:2016gsx}.
It is the different signs in $\bfv\cdot\bfK$ of the inward and outward
neutrino fluxes that produce the crossing in the effective 
spectrum $\tg(\tomega)$ in a stationary neutrino gas when there is no
crossing in the actual neutrino angular distribution. Obviously,
this argument 
does not apply to the temporal flavor instability (i.e.\ with
$\text{Im}(\Omega)\neq 0$) which still requires a crossed neutrino
angular distribution \cite{Izaguirre:2016gsx}.

We have applied the linear flavor stability analysis to a
multi-bulb supernova model. As envisioned by Sawyer fast flavor conversions 
can take place in this model because different neutrino 
species have different neutrino spheres. However, it is somewhat
surprising that it is because of the presence 
of a large matter density that this flavor conversion is possible.
It has been pointed out that the matter effect can be (partially)
``cancelled'' in a non-stationary neutrino gas
\cite{Abbar:2015fwa,Dasgupta:2015iia}. However, a significant
cancellation of the matter effect occurs only for rapidly pulsating
neutrino oscillation modes of frequency $\sim\lambda$ which is
extremely high near or inside
the neutrino sphere. The matter effect discussed in this work is still
relevant to most other neutrino oscillation modes which  are probably
more important  in the real supernova environment.

Of course, our result about supernova neutrinos is only suggestive
because of the various 
assumptions of the multi-bulb model. A real supernova is dynamic and
does not have clear-cut neutrino spheres. There also exist neutrinos
propagating inward which can affect neutrino flavor conversions
\cite{Cherry:2012zw,Izaguirre:2016gsx}. A more definite conclusion
requires simulations 
of neutrino flavor oscillations in time-dependent, multi-dimensional
supernova models with accurate neutrino angular distributions.

\section*{Acknowledgments}
  We thank L.~Ma, J.~Martin, S.~Shalgar, and Y.-Z.~Qian
  for useful discussions and G.~Raffelt for reading the manuscript and
  providing valuable suggestions. We also thank T.~Janka for providing various
  supernova models.  H.D.\ would like to thank the hospitality of
  YITP, Kyoto
  where this work was started and the nuclear physics group at UMN
  where it was finished.
  This work was supported by DOE EPSCoR grant No.\ DE-SC0008142 and DOE
  NP grant No.\ DE-SC0017803 at UNM.

\appendix
\section{Axially symmetric fast oscillations in an outward
  flowing neutrino  flux  
\label{sec:appendix}}
To find out the criterion for the occurrence of the axially symmetric
 fast flavor conversions in an outward flowing neutrino flux 
(Sec.~\ref{sec:outflux}), we again assume that there exists a flavor
instability for $\mu$ within $(\mu_-,\mu_+)$. As
$\mu\rightarrow \mu_- +0^+$, $\Omega\rightarrow\gamma_\rmc$, and
\begin{align}
I[v_z^m] \longrightarrow \frac{\mu}{2}\left[P[v_z^m] +
\rmi\pi G(v_\rmc) \frac{v_\rmc^{m+1}}{\bar\lambda}\right],
\end{align}
where 
\begin{align}
v_\rmc &= \frac{\bar\lambda}{\Phi_z-\gamma_\rmc},\\
P[f(v_z)] &= \mathcal{P}\!\int_0^1\frac{G(v_z) f(v_z)\,\rmd v_z}
{(\gamma_\rmc-\Phi_z)v_z +\bar\lambda}.
\end{align}
At this limit, the imaginary part of Eq.~\eqref{eq:Ia} can be
written as
\begin{align}
\left(P[(v_z-v_\rmc)^2]+\frac{1-v_\rmc^2}{\mu_-/2}\right) G(v_\rmc) v_\rmc = 0,
\end{align}
which is possible if $G(v_\rmc)=0$. 

If $G(v_z)$ has only one simple crossing point (at $v_\rmc$) within the range
of $[0,1]$, then function
\[\frac{G(v_z)}{(\gamma_\rmc-\Phi_z)v_z+\bar\lambda}
= \frac{1}{\gamma_\rmc-\Phi_z} \frac{G(v_z)}{v_z-v_\rmc}\]
has no crossing within the same range. In this case, we can define
\begin{align}
\avg{f(v_z)} = \frac{P[f(v_z)]}{P[1]}
\end{align}
as an average of $f(v_z)$ within $[0,1]$, and the real part of
Eq.~\eqref{eq:Ia} implies that
\begin{align}
\frac{4/\mu_-}{P[1]} =
\left(1-\avg{v_z^2}\right)\pm \sqrt{\left(1-\avg{v_z^2}\right)^2 +
  4\left(\avg{v_z^2}-\avg{v_z}^2\right)}.
\end{align}
The two solutions in the above equation require opposite crossing directions of
$G(v_z)$ at $v_\rmc$. 

Therefore, a condition
for the axially symmetric flavor
oscillations to occur in an outward flowing neutrino flux is that its 
angular distribution $G(v_z)$ has a crossing point within the
range of $[0,1]$, but the 
sign of this crossing is not constrained.

 \bibliographystyle{elsarticle-num}
\bibliography{crospec}

\end{document}